\documentclass{PoS}

\newcommand{\rhocrit}{\rho_c}
\newcommand{\rhorms}{\rho_{\rm rms}}

\newcommand{\cvir}{c_{\rm vir}}
\newcommand{\Rvir}{R_{\rm vir}}
\newcommand{\Mvir}{M_{\rm vir}}
\newcommand{\rs}{r_{\rm s}}

\newcommand{\Nvir}{N_{\rm vir}}
\newcommand{\Ms}{M_{\rm s}}
\newcommand{\cM}{$\rm c_{\rm vir}-M_{\rm vir}$~}
\newcommand{\rhos}{\rho_{\rm s}}

\newcommand{\mnras}{MNRAS}
\newcommand{\apj}{ApJ}
\newcommand{\apjl}{ApJ}
\newcommand{\apjs}{ApJS}

\def \LCDM {\ifmmode \Lambda{\rm CDM} \else $\Lambda{\rm CDM}$ \fi}
\def \kms {\ifmmode  \,\rm km\,s^{-1} \else $\,\rm km\,s^{-1}  $ \fi }
\def \kpc {\ifmmode  {\rm kpc}  \else ${\rm  kpc}$ \fi  }  
\def \Msun {\ifmmode M_{\odot} \else $M_{\odot}$ \fi} 
\def \hMsun {\ifmmode h^{-1}\,\rm M_{\odot} \else $h^{-1}\,\rm M_{\odot}$ \fi}

\title{The Redshift Evolution of $\Lambda$CDM Halo Parameters}

\ShortTitle{The Redshift Evolution of $\Lambda$CDM Halo Parameters}

\author{\speaker{Juan Carlos Mu\~noz-Cuartas}$^a$, Andrea Macci{\`o}$^b$, Stefan Gottl\"{o}ber$^a$, Aaron Dutton$^c$%
  \thanks{CITA National Fellow.}\\
  $^a$Astrophysikalisches Institut Potsdam\\
  $^b$Max Planck Institut F\"{u}r Astronomie\\
  $^c$Department of Physics and Astronomy, University of Victoria\\
  E-mail: \email{jcmunoz@aip.de}}

\abstract{We study the mass and redshift dependence of the
  concentration parameter in Nbody simulations spanning masses from
  $10^{10} \hMsun$ to $10^{15} \hMsun$ and redshifts from 0 to 2. We
  present a series of fitting formulas that accurately describe the
  time evolution of the concentration-mass relation since z=2. Using
  arguments based on the spherical collapse model we study the
  behaviour of the scale length of the density profile during the
  assembly history of haloes, obtaining physical insights on the
  origin of the observed time evolution of the concentration mass
  relation. We present preliminary results of the implementation of
  this model in the prediction of the values of the concentration
  parameter for different masses and redshifts.}

\FullConference{Cosmic Radiation Fields: Sources in the early Universe\\
                November 9-12, 2010\\
                Desy, Germany}

\begin{document}

\section{Introduction}

The current standard cosmological paradigm, the so called $\Lambda$CDM
universe, establishes that the universe has a spatially flat geometry
whose dynamics is nowadays dominated by dark energy with a minor
contribution from the dark matter, both of them are of unknown nature
and where neither radiation nor baryonic matter play an important role
in the present dynamical state of the universe. In this scenario,
structures grow as gravitational instabilities in the dark matter
density field where the first objects are the smaller ones and
subsequent mergers and accretion of the small objects on to the big
ones lead to a hierarchical scenario of growth of structures. Galaxies
are supposed to form as gas accretes and cools down in the
gravitational potential well of those previously formed dark matter
haloes \cite{WhiteRees78}. Dark matter haloes play a crucial role in
the study of the formation of galaxies and the evolution of the
universe, they provide the environment where baryons can cool down to
form stellar systems, which finally are the objects we can observe in
the universe and are the ones responsables for much of the cosmic
radiation field we can account for.

The properties of dark matter haloes can be characterized with several
different parameters, one of the most important ones is the
concentration parameter which accounts for the shape of the mass
density profile and therefore has important implications in
determining the properties of the galaxies forming inside
\cite{MoMaoWhite}, as well as it has direct influence in the flux of
radiation produced during the annihilation process of dark matter in
the center of the halo \cite{Hutsi2009}. Previous works have been
addressed in the study of the structure of dark matter
haloes. \cite{NFW97} (NFW) proposed that the characteristic density of
dark matter haloes was directly proportional to the density of the
universe at time of formation, making possible to connect today
properties of the dark matter density profile to the halo formation
history and to the evolution of the expanding
universe. \cite{Bullock2001a} and \cite{Bullock2001b} have studied the
mass dependence of the concentration parameter, they found a power law
mass dependence of the concentration parameter that scales in time
with scale factor. \cite{Wechsler2002} found a strong correlation
between the concentration parameter and the mass accretion history of
haloes, and confirmed the claims of NFW and \cite{Bullock2001a} where
the concentration of the halo is related to their definition of time
of formation of the halo. \cite{ENS2001} investigated the power
spectrum dependence of the concentration
parameter. \cite{zhao2003a},\cite{zhao2003b},\cite{zhao2009}
re-addressed the problem of the properties of the halo mass
distribution and studied the connection between the mass accretion
history (MAH) and the concentration parameter using a large suite of
simulations. They found a relation between the scale length of the
halo $\rs$ and the mass interior to that radius $\Ms$. Using that
relation in his model for the MAH and an appropriated choice of the
time of transition between the two different modes of accretion
enabled them to model the mass and redshift dependence of the
concentration parameter. \cite{Maccio07} and \cite{Maccio08} have
studied the cosmology and mass dependence of the concentration, shape
and spin parameters at z=0 and revised the models of NFW and
\cite{Bullock2001a}. In \cite{Munoz-Cuartas2010} we studied the mass
and redshift dependence of spin, shape and concentration parameters
for a WMAP5 cosmology and found the evolution of the concentration
parameter to be due to the evolution of the inner halo mass
distribution (inside $\rs$) and suggest that this evolution can be
modeled as a spherical perturbation growing in the inner region of the
halo. This description gives physical insight in to the understanding
of the mass and redshift dependence of the concentration parameter.

In this work we present in more detail those results presented in
\cite{Munoz-Cuartas2010} and focus on the redshift dependence of the
concentration parameter. We show our advances in the implementation of
those ideas in the construction of a model able to make predictions
for the values of the concentration parameter for different masses and
redshifts, this may be important for the modeling of the properties of
galaxies as well as for the study of sources of annihilation of dark
matter in the early universe.

\section{Simulations and methods}

\subsection{Simulations}

All simulations in this work have been performed with {\sc pkdgrav}, a
tree code written by Joachim Stadel and Thomas Quinn \cite{Stadel2001PhDT}. The code uses spline kernel softening, for which the forces
become completely Newtonian at 2 softening lengths.  Individual time
steps for each particle are chosen proportional to the square root of
the softening length, $\epsilon$, over the acceleration, $a$: $\Delta
t_i = \eta\sqrt{\epsilon/a_i}$. Throughout, we set $\eta = 0.2$, and
we keep the value of the softening length constant in comoving
coordinates during each run. The physical values of $\epsilon$ at
$z=0$ are listed in Table \ref{tab:sims}.  Forces are computed using
terms up to hexadecapole order and a node-opening angle $\theta$ which
we change from $0.55$ initially to $0.7$ at $z=2$.  This allows a
higher force accuracy when the mass distribution is nearly smooth and
the relative force errors can be large.  The initial conditions are
generated with the {\sc grafic2} package (\cite{Bertschinger2001}).  The
starting redshifts $z_i$ are set to the time when the standard
deviation of the smallest density fluctuations resolved within the
simulation box reaches $0.2$ (the smallest scale resolved within the
initial conditions is defined as twice the intra-particle distance).

We have set the cosmological parameters according to the fifth-year
results of the Wilkinson Microwave Anisotropy Probe mission WMAP5
\cite{Komatsu2009}, namely, $\Omega_m = 0.258$, $\Omega_L = 0.742$,
$n=0.963$, $h = 0.72$, and $\sigma_8 = 0.796$, where $\Omega_m$ and
$\Omega_L$ are the values of the density parameters at z=0. Table
\ref{tab:sims} lists all of the simulations used in this work.  We
have run simulations for several different box sizes, which allows us
to probe halo masses covering the entire range $10^{10}$ \hMsun $< M <
10^{15} h^{-1}\,\rm M_{\odot}$.  In addition, in some cases we have
run multiple simulations for the same cosmology and box size, in order
to test for the impact of cosmic variance (and to increase the final
number of dark matter haloes).


\begin{table}
\begin{center}
\begin{tabular}{|c|c|c|c|c|c|}\hline \hline

Name  & Box Size & N & $m_p$ & $\epsilon$ & $N_{min}> 500$  \\
      &          &   &       &            &  z=0,2          \\ \hline \hline

B20   & 14.4   & $250^3$  & 1.37e7  & 0.43 & 974,   1006  \\
B30   & 21.6   & $300^3$  & 2.68e7  & 0.64 & 1515,  1399  \\
B40   & 28.8   & $250^3$  & 1.10e8  & 0.85 & 1119,  993   \\
B90   & 64.8   & $600^3$  & 9.04e7  & 0.85 & 13587, 12177 \\
B180  & 129.6  & $300^3$  & 5.78e9  & 3.83 & 2300,  510  \\
B300  & 216.0  & $400^3$  & 1.13e10 & 4.74 & 5840,  707  \\
$\rm{B300_2}$  & 216.0  & $400^3$  & 1.13e10 & 4.74 & 5720, 766 \\
\hline
\hline
\end{tabular}
\end{center}
\caption{Table of simulations used in this work. Note that the name of
  the simulation is related to the box size in units of Mpc. $N$
  represents the number of total particles in the box. $\epsilon$
  represents the force softening length in units of kpc $h^{-1}$ and
  the last column gives the number of haloes with more than 500
  particles at $z=0$ and $z=2$.  Masses of particles are in units of
  \hMsun\ and box sizes in units of Mpc $h^{-1}$, with $h=0.72$.}
\label{tab:sims}
\end{table}


\subsection{Halo properties: Concentrations}

In this work we identify dark matter halos using a spherical
overdensity (SO) algorithm and use a time varying virial density
contrast determined using the fitting formula presented in
\cite{ByN98}. We include in the halo catalogue all the haloes with
more than 500 particles inside the virial radius ($N_{\rm vir}>500$).

To compute the concentration of a halo we first determine its density
profile.  The halo centre is defined as the location of the most bound
halo particle (we define the most bound particle as the particle with
the lowest potential energy, no care about binding energy is taken
here), and we compute the density ($\rho_i$) in 50 spherical shells,
spaced equally in logarithmic radius.  Errors on the density are
computed from the Poisson noise due to the finite number of particles
in each mass shell.  The resulting density profile is fitted with a
NFW profile:

\begin{equation}
  \frac{\rho(r)}{\rhocrit} = \frac{\delta_{\rm c}}{(r/\rs)(1+r/\rs)^2},
  \label{eq:nfw}
\end{equation}

During the fitting procedure we treat both $\rs$ and $\delta_c$ as free
parameters.  Their values, and associated uncertainties, are obtained via a
$\chi^2$ minimization procedure using the Levenberg \& Marquardt method.  We
define the r.m.s. of the fit as:
\begin{equation}
\rhorms = \frac{1}{N}\sum_i^N { (\ln \rho_i - \ln \rho_{\rm m})^2}
\label{eq:rms}
\end{equation}
where $\rho_{\rm m}$ is the fitted NFW density distribution.
Finally, we define the concentration of the halo, $\cvir
\equiv\Rvir/\rs$, using the virial radius obtained from the SO
algorithm, and we define the error on $\log c$ as
$(\sigma_{\rs}/\rs)/\ln(10)$, where $\sigma_{\rs}$ is the fitting
uncertainty on $\rs$.

\section{Results}

\begin{figure}
  \begin{center}
    \includegraphics[width=7.0cm,angle=270]{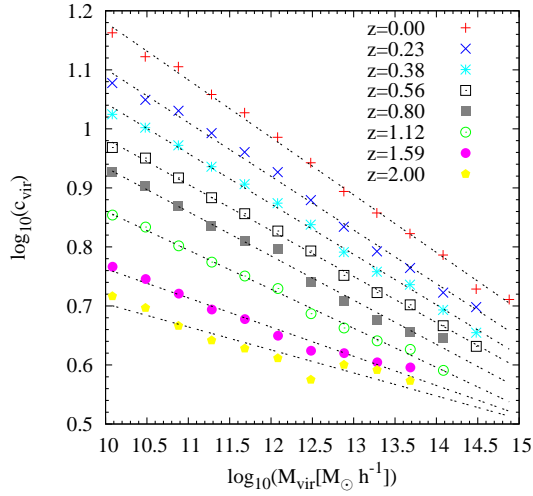}
  \end{center}
  \caption{Mass and redshift dependence of the concentration
    parameter. The points show the median of the concentration as
    computed from the simulations, averaged for each mass bin. Lines
    show their respective linear fitting to eq. 3.1. }
  \label{fig:Cwithfit}
\end{figure}

In figure \ref{fig:Cwithfit} we show the median \cM relation for
relaxed haloes in our sample at different redshifts.  Haloes have been
binned in mass bins of 0.4 dex width, the median concentration in each
bin has been computed taking into account the error associated to the
concentration value (see \ref{ssec:c}, and M08).  In our mass range
the \cM relation is well fitted by a single power law at almost all
redshifts. Only for $z=2$ we see an indication that the linearity of
the relation in log space seems to break, in agreement with recent
findings by \cite{Klypin2010}.

The best fitting power law can be written as:
\begin{equation}
\log(c) = a(z)\log(M_\mathrm{vir}/[\hMsun]) + b(z)
\label{eq:Cfit}
\end{equation}

The fitting parameters $a(z)$ and $b(z)$ are functions of redshift,
the evolution of $a$ and $b$ can be itself fitted with two simple
formulas that allow to reconstruct the \cM relation at any redshifts:

\begin{equation}
a(z) = wz  - m
\label{Pfita}
\end{equation}
\begin{equation}
b(z) = \frac{\alpha}{(z+\gamma)}+\frac{\beta}{(z+\gamma)^2}
\label{Pfitb}
\end{equation}

\noindent
Where the additional fitting parameters have been set equal to:
$w=0.029$, $m=0.097$, $\alpha=-110.001$, $\beta=2469.720$ and
$\gamma=16.885$.  This double fitting formulas are able to recover the
original values of the halo concentration with a precision of 5\%, for
the whole range of masses and redshifts inspected. It has been shown
by \cite{Trenti2010} that using $\Nvir$ between 100 and 400
particles is enough to get good estimates for the properties of halos,
nevertheless in order to look for systematics we re-computed $\cvir$
varying the minimum number of particle inside $R_{\rm vir}$, using
200, 500 and 1000 particles. No appreciable differences (less than
2\%) were found in our results for the median.


As can be seen from our results, the mass and redshift dependence of
the concentration parameter is considerably different from a simple
scaling with redshift \cite{Bullock2001a}. As has been already noted
by \cite{Wechsler2002} and \cite{zhao2003a},\cite{zhao2009} the
evolution of the concentration parameter is strongly coupled to the
growth history of the halo, we will follow this idea to study the
physical mechanism behind its evolution. To do so, we built merger
trees for all haloes at z=0 in our simulations. This time we used
haloes with $\Nvir > 200$ to be able to follow the evolution of haloes
up to earlier times. As was already mentioned, using 200 particles
inside the virial radius still gives good results in the estimation of
the properties of the halo we are interested in and for simplicity we
will only present results from our box B90.

Our methodology is simple, since we define the concentration parameter
$\cvir$ as the ratio between the virial radius $\Rvir$ and the scale
length radius $\rs$, we study the evolution of $\cvir$ tracking the
evolution of these quantities along the merger tree of haloes. It is
reasonable to think that the properties of halos we observe at z=0
(and at any redshift) are inherited from the most massive progenitor,
so we only trace back the properties of the halo along this branch of
the tree.

\begin{figure}
  \includegraphics[width=7.0cm,angle=270]{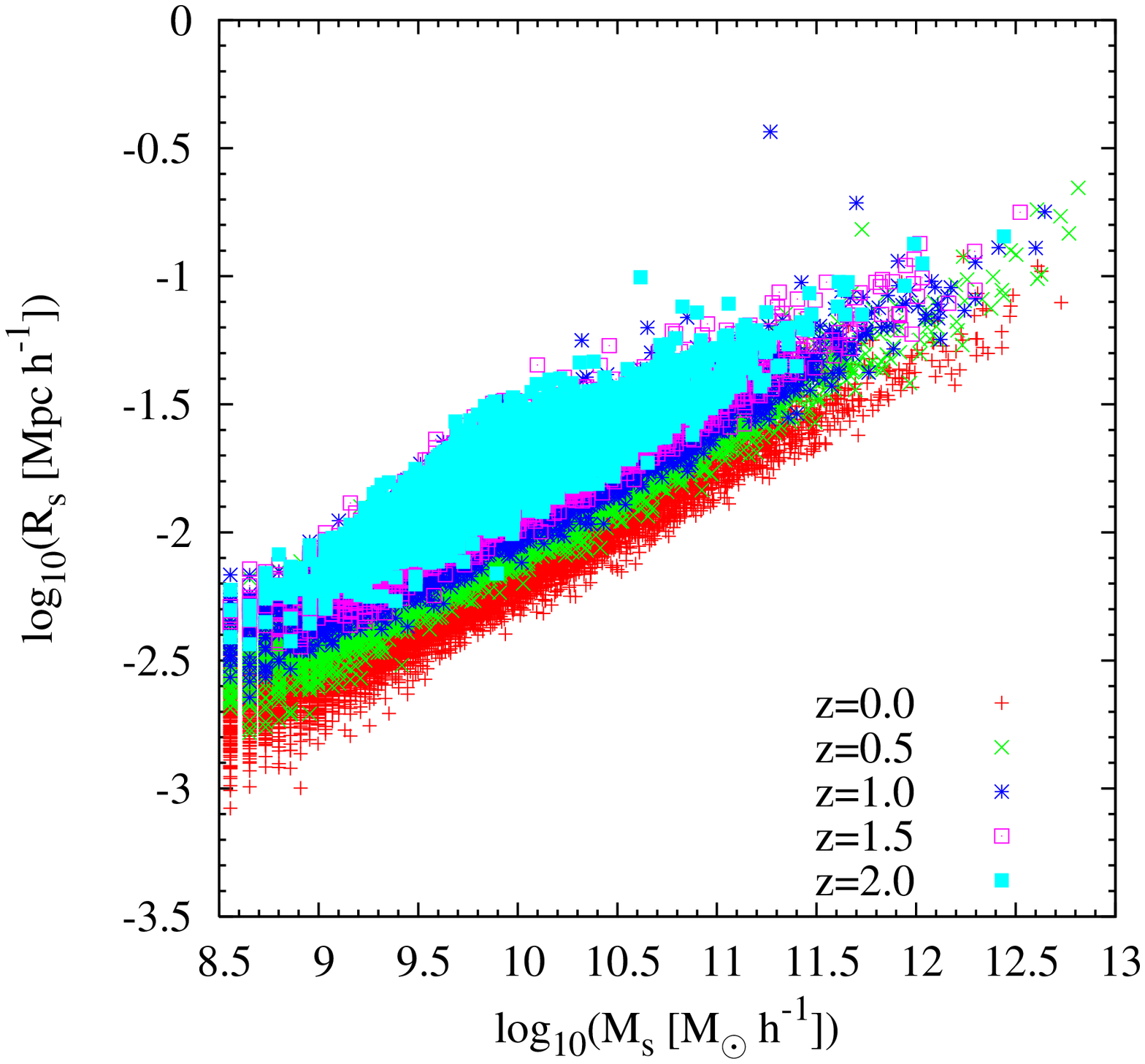}
  \includegraphics[width=7.0cm,angle=270]{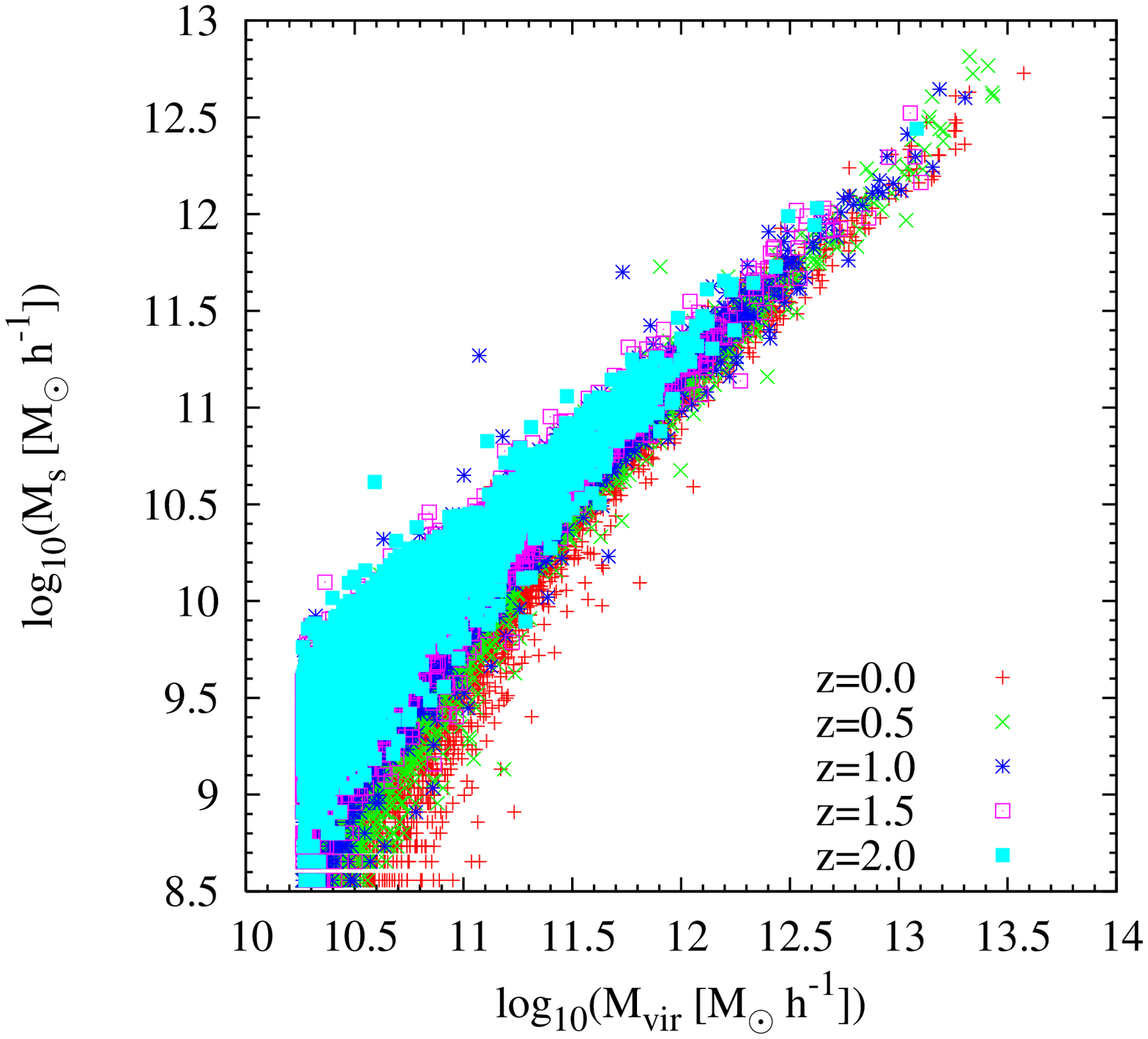}
  \caption{(Left) Relation between $\Ms$ and $\rs$ for haloes at
    different redshift. For all redshifts the mean slope equals 0.38
    with very low scatter, while the normalization is clearly redshift
    dependent. The scatter in the relation is mostly due to the
    intrinsic scatter in the values of $\rs$. (Right) Relation between
    $\Mvir$ and $\Ms$. Clearly both quantities are directly related
    with a weak redshift dependence in the normalization.}
  \label{fig:MsRs}
\end{figure}

First we explore the relation between $\Ms$, defined as the mass
inside $\rs$, and $\rs$. Figure \ref{fig:MsRs} shows clearly the
relation between both quantities, if $\rhos(z)$ is the mean mass
density inside $\rs$ at a given redshift $z$, then

\begin{equation}
r_s(z) = \left( \frac{3\Ms(z)}{4\pi\rhos(z)} \right)^{\alpha}
\label{eq:rsdef}
\end{equation}

if $\rhos(z)$ where set to be mass independent one could verify that
the power index $\alpha$ in \ref{eq:rsdef} equals exactly 1/3, any
deviation from that value would be due to a mass dependence on
$\rhos(z)$. The analysis on our data shows robustly that $\alpha
\approx 0.38$ close but clearly different from 1/3, which may imply
that $\rhos(z)$ has a weak dependence on $\Ms$. This behavior can be
compared with the one expected from the relation between $\Rvir$ and
$\Mvir$ where the power index of the equivalent relation
\ref{eq:rsdef} is exactly 1/3 without any mass dependence on the mean
density of the halo, which is constrained by the values of the
overdensity contrast predicted by spherical collapse and the critical
density of the universe. Finally, Figure \ref{fig:MsRs} also shows
that $\Ms$ and $\Mvir$ are related, so, any dependence of $\rhos(z)$
with $\Ms$ will translate in to a dependence with $\Mvir$.

\begin{figure}
  \includegraphics[width=7.0cm,angle=270]{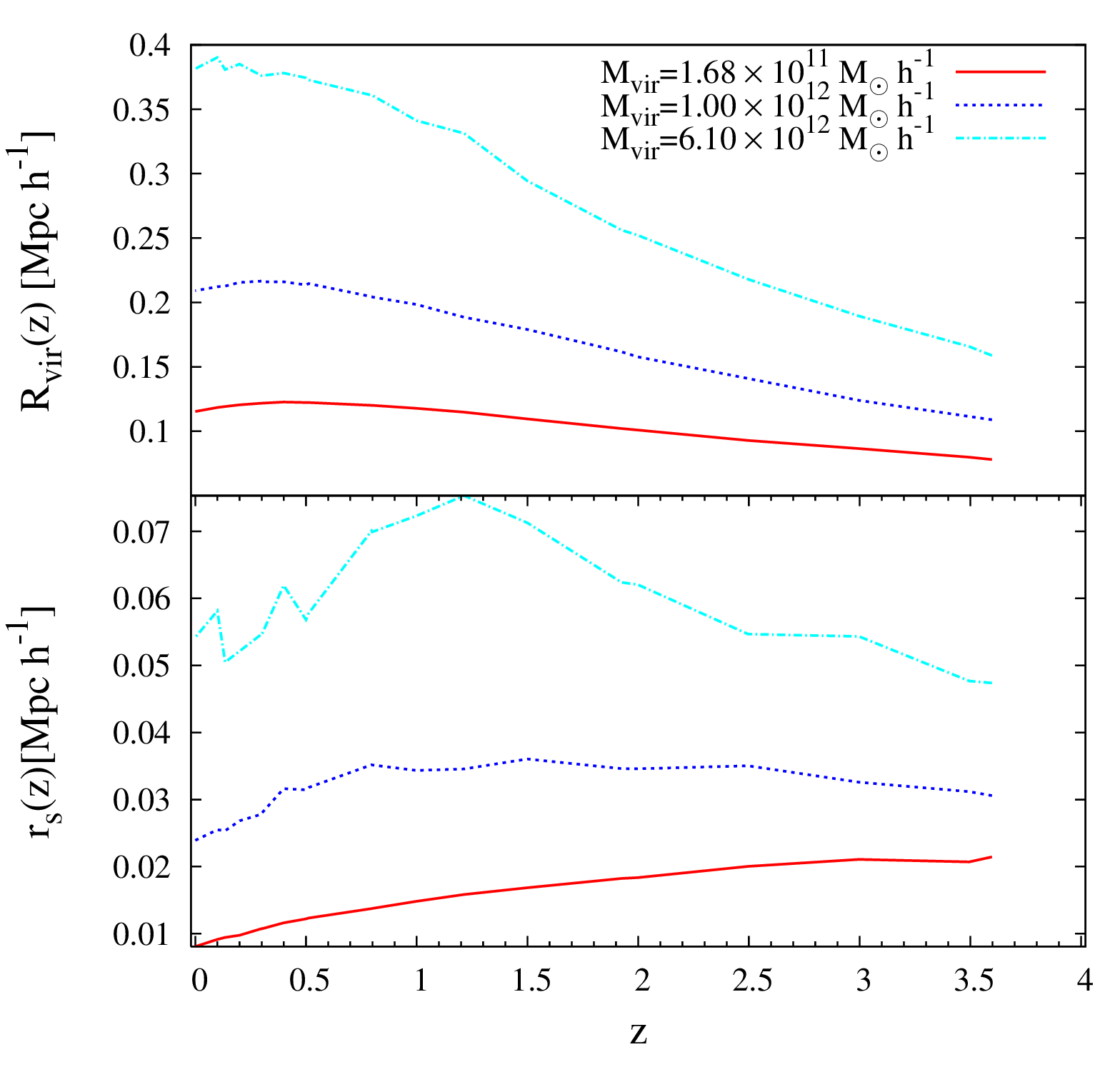}
  \includegraphics[width=7.0cm,angle=270]{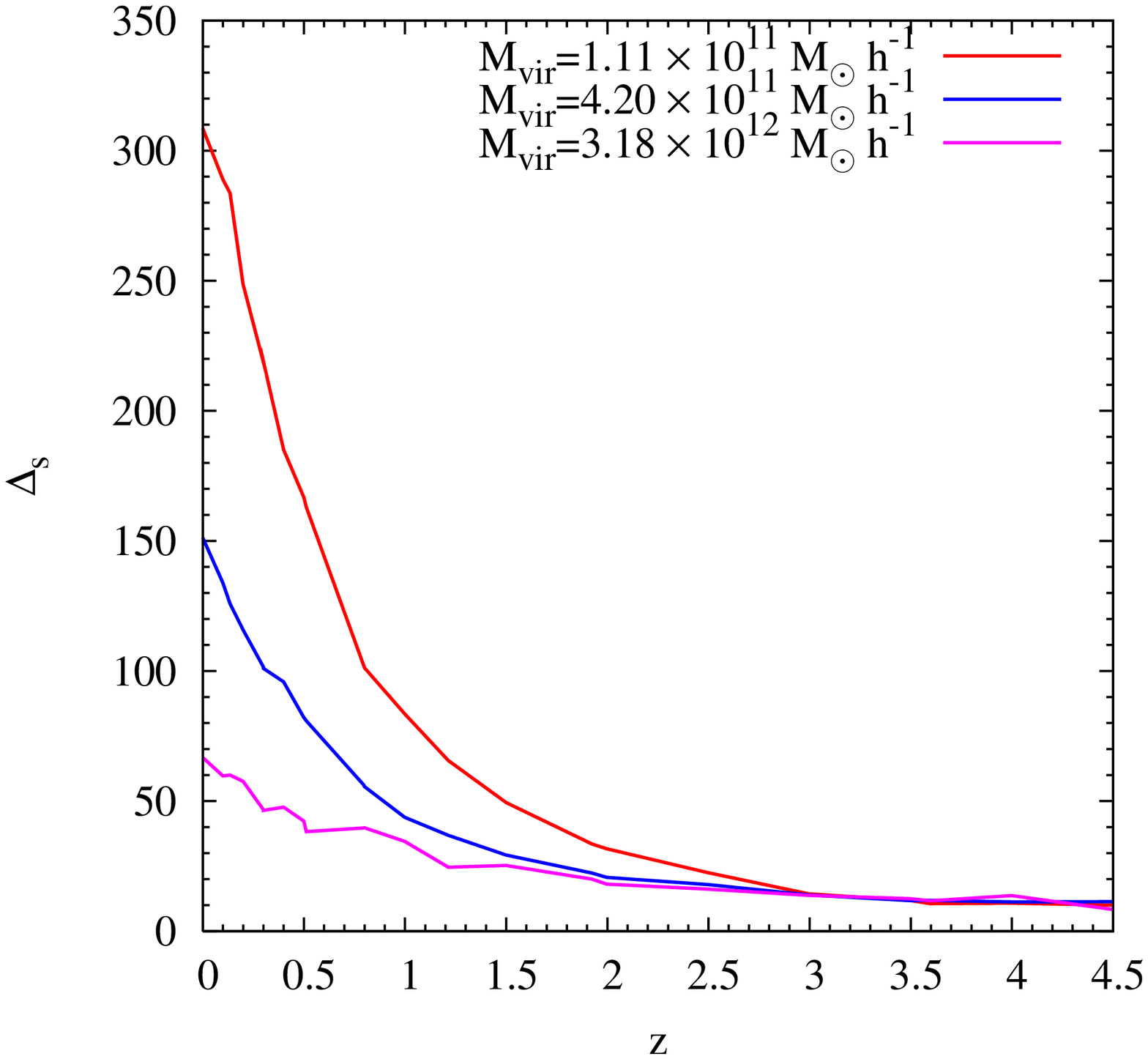}
  \caption{(Left-Top) Time evolution of the virial radius as computed
    from the averaged merger histories for haloes with final mass of
    $1.68\times10^{11}$, $1.0\times10^{12}$ and $6.1\times10^{12}
    h^{-1}\,\rm M_{\odot}$. (Left-Bottom) Time evolution of the
    averaged scale length $\rs$ of the dark matter haloes in the same
    mass bins. (Right) $\Delta_{\rm s}(z)$ as a function of redshift
    for three different mass bins computed as the ratio
    $\rhos(z)/\rho_{vir}(z)$ along the merger tree.}
  \label{fig:Rvir}
\end{figure}

This similarity in the relations between $\rs$ with $\Ms$ and $\Rvir$
with $\Mvir$ suggest that one may think in an analogous treatment for
$\rs$ and $\Ms$ as is done for $\Rvir$ and $\Mvir$. Such an analogous
analysis is supported by the results of the time evolution of the
radial scale length $\rs$ and virial radius $\Rvir$ along the merger
trees. Figure \ref{fig:Rvir} shows the time (redshift) evolution of
$\Rvir$ and $\rs$. As it can be seen in the figure, the behavior of
both quantities show a similar trend. They grow with decreasing
redshift, reach a maximum and then start to decrease. The time at
which the maximum is reached depends on the final virial mass of the
halo, being the low mass ones the firsts reaching that point.

The similarities observed in figures \ref{fig:MsRs} and \ref{fig:Rvir}
suggest that one can describe the evolution of the inner region of the
halo (the one enclosed by $\rs$) in a similar way as done for the
outer halo. We assume the inner region of the halo to be a
perturbation of mean density $\rhos(z)$ that evolves within the
background of mean density $\rho_{vir}(z) =
\Delta_{vir}(z)\rho_c(z)$. In analogy with the spherical collapse
model we want to look for the evolution of the density contrast of
this perturbation: $\Delta_s(z) = \rhos(z)/\rho_{vir}(z)$. This inner
density contrast is well described by the following formula:

\begin{equation}
  \Delta_{\rm s} (z)= \frac{A}{z+\epsilon(M)}
  \label{eq:rsth}
\end{equation}

where $A=50$ and $\epsilon(M) = 0.3975\log(M_{\rm vir}(z=0)/[\hMsun])-
4.312$ best reproduce our data. Equation \ref{eq:rsth} implies that:
i) $\rho_{\rm s} > \rho_{\rm vir}$ at all redshifts, ii) $\Delta_{\rm
  s}$ is a growing function of the redshift, implying a fast growth of
the inner density with respect to the mean density of the halo and
iii) $\Delta_{\rm s}$ depends on the final mass of the halo, and it
has lower values for high mass haloes. This mass dependence of
$\Delta_{\rm s}$ is also justified from the analysis of
figure \ref{fig:MsRs} where it was shown that the mean density of
the inner region of the halo must be mass dependent.

\begin{figure}
  \begin{center}
    \includegraphics[width=7.0cm,angle=270]{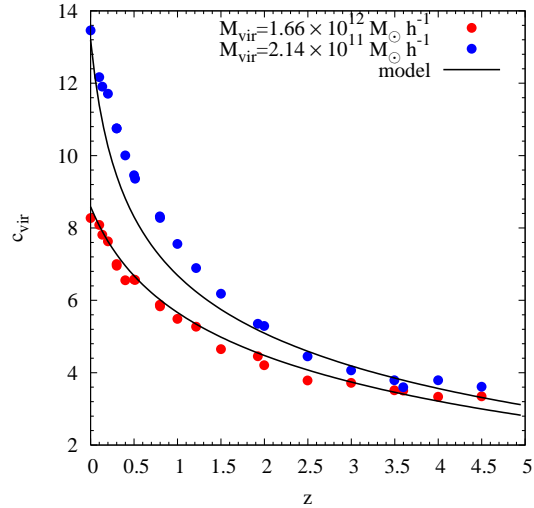}
  \end{center}
  \caption{Data and model of the redshift dependence of the
    concentration parameter. Data points are computed as the ratio
    between the mean values of $\Rvir$ and $\rs$ in the respective
    mass bins along the branch of the most massive progenitors of the
    tree, the solid lines are the prediction of the model.}
  \label{fig:Cvirmodel}
\end{figure}

One can use all of the previous results to try to reproduce the mean
values of the concentration parameter as a function of time for
different halo masses. Figure \ref{fig:Cvirmodel} shows the result of
modeling $\cvir$ after modeling $\Rvir$ and $\rs$ with the use of eq
\ref{eq:rsdef}. For that $\Mvir(z)$ was modeled as a damped
exponential law (\cite{McBride2010}, \cite{Munoz-Cuartas2010}) and
$\Ms$ was assumed to follow a power law with $\Mvir$ with appropriated
values for the power law index and normalization.

Although the model works quite well in describing the time dependence
of the concentration parameter, it is clear that it works much better
for the high mass than for the low mass regime, although for the low
mass haloes the model fits very well at high redshift. It is clear
that the model performs better for halos in regimes where the
nonlinear effects still are not so strong to be predictable by the way
spherical collapse approach is implemented in this work.

\section{Conclusions}

We present results of the study of the mass and redshift dependence of
the concentration parameter of dark matter haloes. In our mass and
redshift range the \cM relation always follows a power law behavior.
We confirmed that the redshift dependence of such relation is more
complex than a simple $(1+z)^{-1}$ scaling as proposed by
\cite{Bullock2001a}, with both the normalization and the slope of the
relation changing with cosmic time.  We also found that for increasing
redshifts ($z \approx 2$) the power law behavior seems to break, in
agreement with recent studies (e.g. \cite{Klypin2010}).  Thanks to our
multiple box simulations we tested our results against resolution
effects and find them to be stable once a sufficient large number of
particles is used $N_{\rm vir}>500$.

In order to improve our understanding on the redshift evolution of the
\cM relation we look at the individual evolution with time of $\rs$
and $\Rvir$.  Both these length scales grow with decreasing redshift
until a maximum is reached, then they start to decrease towards
$z=0$. There is a clear analogy between the collapse of a linear
perturbation and the behavior of $\rs$ and $\Rvir$. We found that we
can model the evolution of the inner part of the halo as a decoupled
spherical perturbation growing inside the central region of the
halo. The temporal offset between the ``turning points'' of the
perturbations associated with $\rs$ and $\Rvir$ is able to explain the
observed redshift evolution of the \cM relation. Using this model we
presented first results showing that the model we propose may be used
to predict the evolution of the halo mass density profile for
arbitrary mass and redshifts, which will have important applications
in the modeling of the mass distribution of halos in redshift or mass
regimes where simulations have limited resolution.

\end{document}